\documentclass[prl,twocolumn,superscriptaddress,showpacs]{revtex4}

\usepackage{graphicx}
\usepackage{amsmath}
\usepackage{amssymb}

\begin{document}

\newcommand{\ud}{{\mathrm d}}
\newcommand{\sech}{\mathrm{sech}}

\newcommand{\bs} {\boldsymbol}

\newcommand{\diff} {\mathrm{d}}
\newcommand{\f} {\frac}

\newcommand{\ga} {\alpha}
\newcommand{\gb} {\beta}
\newcommand{\gc} {\gamma}

\newcommand{\MB} {\mathrm{M}}
\newcommand{\J}  {\mathrm{J}}
\newcommand{\MJ}  {\mathrm{H}}

\newcommand{\Obs}  {\mathrm{O}}

\newcommand{\tot}  {\mathrm{tot}}
\newcommand{\Z}  {\mathcal{Z}}
\newcommand{\kB} {k_\mathrm{B}}


\title{Nucleation of breathers via stochastic resonance in nonlinear lattices}


\author{David Cubero}
\affiliation{Departamento de F\'{\i}sica Aplicada I, EUP, Universidad de Sevilla, Calle Virgen de \'Africa 7, 41011 Sevilla, Spain}
\author{Jes\'us Cuevas}
\affiliation{Departamento de F\'{\i}sica Aplicada I, EUP, Universidad de Sevilla, Calle Virgen de \'Africa 7, 41011 Sevilla, Spain}
\author{P. G. Kevrekidis}
\affiliation{Department of Mathematics and Statistics, University of Massachusetts, Amherst, Massachusetts 01003-4515, USA}

\date{\today}

\begin{abstract}
By applying a staggered driving force in a prototypical discrete model
with a quartic nonlinearity, we demonstrate the spontaneous formation and destruction of discrete
breathers with a selected frequency due to thermal fluctuations. The phenomenon exhibits the striking features of stochastic resonance (SR): a nonmonotonic behavior as noise is increased and breather generation under subthreshold conditions. The corresponding peak is associated with a matching between the external driving frequency and the breather frequency at the average energy given by ambient temperature. 
\end{abstract}




\maketitle

Intrinsic localized modes, often referred to as discrete 
breathers, have been the focus of a number of theoretical studies 
(for a recent review see e.g. \cite{FG08}). One of the main reasons 
for the large interest in these modes is the fact that they provide a 
natural setup for energy localization, a  paradigm of interest to many 
areas of physics as well as chemistry and biology.
Such discrete breathers have been rigorously proven to exist as 
time-periodic localized excitations in nonlinear Hamiltonian 
lattices \cite{macaub94}. Ever since, they have been experimentally
observed in a wide variety of different media, ranging from 
optical waveguides and photorefractive crystals to micromechanical
cantilever arrays and Josephson junctions, as well as in Bose-Einstein
condensates and layered antiferromagnets, among many others 
\cite{cfk04,FG08}.

In the present work, having in mind realistic physical systems, 
we will focus on the nucleation of such breathers in a prototypical 
model system in which friction and ambient noise are also present. 
This will lead us to 
the consideration of stochastic resonance (SR), the counter-intuitive phenomenon by which an appropriate 
dose of noise, instead of degradation, produces an enhancement of 
sensitivity of a nonlinear system to external forcing \cite{gamhan98}. In spatially extended systems, SR has been shown for bistable \cite{margam96} or excitable media \cite{junmay95}, usually associated with pattern formation \cite{vilrub97}, see e.g. the review of \cite{gamhan98}.

Our prototypical nonlinear model will consist of a quartic potential
(the so-called hard $\phi^4$ lattice \cite{cheaub96}) with the following 
equation of motion for each oscillator $x_n$ ($n=1,\ldots,N$) with mass $m$
\begin{equation}
 m\ddot{x}_n=-U'(x_n)+k(x_{n+1}+x_{n-1}-2x_n)-\alpha\dot{x}_n+\xi_n+F_n(t),
\label{eq:system}
\end{equation}
where the overdot and the prime denote time and space derivative, respectively; $k$ is the coupling parameter between oscillators; $\alpha$ the damping constant; $F_n(t)$ denotes external driving; and $\xi_n$ is a Gaussian white noise with zero mean and autocorrelation $\langle \xi_n(t)\xi_m(t')\rangle=2D\delta_{nm}\delta(t-t')$. The noise strength parameter $D$ obeys the fluctuation-dissipation relation $k_BT=D/\alpha$, where $k_B$ is the Boltzmann constant and $T$ the ambient temperature. Thus, $D$ is an indicator of the level of noise associated with thermal fluctuations.

The on-site potential is given by $U(x)=a x^2/2+b x^4/4$ with $a,b>0$. This is 
a {\it monostable} nonlinear potential which confers interesting properties to the lattice. The fact that is monostable implies that the noiseless lattice has a single stable steady state, allowing us to focus on the behavior of homoclinic orbits, rather than kink-like heretoclinic orbits, like in the bistable $\phi^4$ case ($a<0$). This is one way in which our work differs from the 
standard bistable models of SR, or the interesting recent work of \cite{milkho09}, which 
considers noise-induced transitions between steady states in very short
Fermi-Pasta-Ulam (FPU) type, boundary-driven chains. 
Moreover, this model possesses a phonon band which can be easily determined upon linearization of (\ref{eq:system}) as $\omega_\mathrm{ph}\in(\sqrt{U''(0)/m},\sqrt{[U''(0)+4k]/m})$ \cite{FG08}. Thus, by exciting breathers with a frequency larger than the maximum frequency of the phonon spectrum, 
$\omega_\mathrm{max}=\sqrt{(a+4k)/m}$, 
these localized modes 
are guaranteed not to interact with the phonons, avoiding effects like 
radiation resonances which slowly dissipate away the energy of discrete breathers \cite{johaub00}.

\begin{figure}[t]
\includegraphics[width=7.5cm,angle=0]{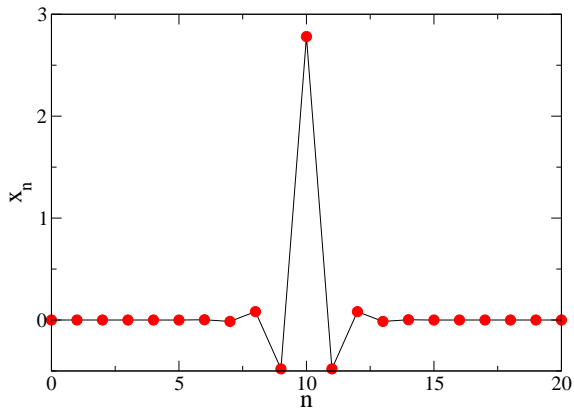}
\caption{
\label{fig:0}
Spatial profile of a staggered high-frequency ($\omega=3$) Hamiltonian breather generated using methods based on the anti-continuous limit \cite{MA96}. The line is a guide to the eye. 
}
\end{figure}

Indeed, the associated Hamiltonian system of this hard model (i.e., Eq.~(\ref{eq:system}) without damping and noise) exhibits exponentially localized in space and periodic in time {\em high} 
(i.e., higher than $\omega_\mathrm{max}$) 
frequency excitations. 
An example is the dynamically stable Sievers-Takeno (ST) \cite{sietak88} 
site-centered mode. As shown in Fig.~\ref{fig:0}, this mode
is {\em staggered}, i.e., the neighboring sites oscillate out of phase. Hence, these breathers will necessitate that we impose staggered periodic driving in order to sustain them in the presence of dissipation. For this reason, we shall consider the following AC driving force
\begin{equation}
 F_n(t)=(-1)^n F_0 \sin(\Omega t).
\label{eq:force}
\end{equation}
Its effect on a deterministic FPU lattice 
has been studied in Refs.~\cite{kholep01,kholep02}. Furthermore, such staggered driving is straightforward
to apply to driven nonlinear electrical lattices \footnote{This can be done
in settings similar to those of Ref.~\cite{satyas07}, by changing the signal and ground leads in every other resistor
leading to a diode in the configuration (L.Q. English, private communication).}. 
Evidently, the driving frequency $\Omega$ must be chosen with the same value as the breather frequency we are interested to excite. 

The equations (\ref{eq:system}) are supplemented 
with (periodic) boundary conditions $x_0=x_N$ and $x_{N+1}=x_1$. Generally, $\phi^4$ models has been commonly used to describe a wide range of 
phenomena from elementary particle collisions and cosmological
domain walls \cite{annoli91} to the behavior of an order parameter near the 
condensation point of a droplet in nucleation theory and phase
transition phenomena \cite{lan67}. They also constitute a 
prototypical nonlinear model in statistical field theory \cite{parisi}.

For the sake of a dimensionless description, we can scale all lengths and times by the characteristic length $\lambda=(a/b)^{1/2}$ and time $\tau=(m/a)^{1/2}$, i.e. $\tilde{x}_n=x_n/\lambda$, $\tilde{t}=t/\tau$, which implies $\tilde{k}=k/a$, $\tilde{\alpha}=\alpha/(am)^{1/2}$, $\tilde{D}=b D/(a^5m)^{1/2}$, and $\tilde{F}_0=F_0 (b/a^3)^{1/2}$. In the following we will omit the tilde symbols for
simplicity. The dimensionless form of the equations of motion (\ref{eq:system}) then reads
\begin{eqnarray}
\ddot{x}_n=-x_n-x_n^3+k(x_{n+1}+x_{n-1}-2x_n)-\alpha\dot{x}_n \nonumber\\
+\xi_n+(-1)^n F_0 \sin(\Omega t).
\label{eq:system:d}
\end{eqnarray}

Let us discuss, for the moment, the deterministic behavior of (\ref{eq:system:d}) (i.e. without noise, $D=0$). As mentioned above, the staggering factor $(-1)^n$ in (\ref{eq:force}) is important in order to sustain the breathers in the presence of dissipation. When that factor is absent, the ST breather shown in Fig.~\ref{fig:0} is always observed to die out, regardless of the driving strength.
However, in addition to this staggering factor, the driving amplitude $F_0$ must be large enough so that the energy supplied by the driving force is able to compensate the energy which is been dissipated by the friction force. Breathers in the dissipative deterministic lattice can be calculated using a method similar to the developed in \cite{MFMF01}. An analysis of those breathers shows that, for $k=1$ and $\alpha=0.1$ and a breather frequency of $\omega=\Omega=3$, there is a threshold at $F_\mathrm{th}=0.6092$. When the driving amplitude is below this threshold (subthreshold amplitude), the dissipation drains away gradually the system's energy and breather modes  cannot be supported. In the following, unless explicitly stated, we will assume $k=1$, $\alpha=0.1$, and $\Omega=3$, which implies $\omega_\mathrm{max}=\sqrt{5}<\Omega$.

\begin{figure}[t]
\includegraphics[width=7.5cm,angle=-90]{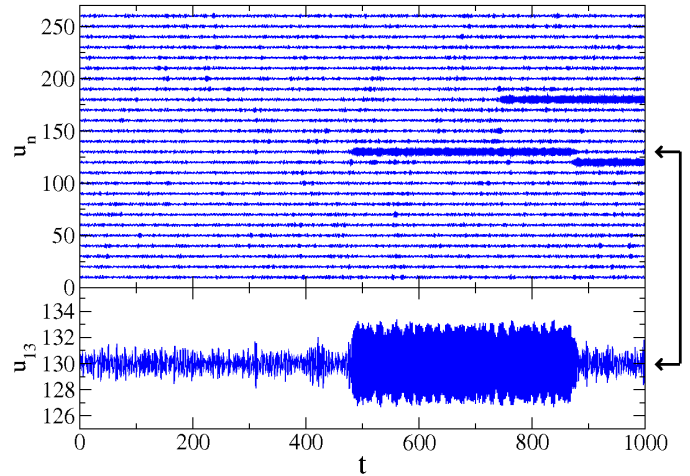}
\caption{
\label{fig:1}
(Color online) Time evolution of the oscillator positions $u_n(t)\!:= x_n(t)+10\,n$ for a system subject to a suprathreshold driving with $F_0=1$ and $\Omega=3$, and noise strength $D=0.1$ starting from a uniform initial condition $x_n(0)=0$. Breathers are observed to form and vanish spontaneously. The bottom panel shows a magnification of the time evolution of an oscillator which becomes the center of a high-frequency breather.
}
\end{figure}

\begin{figure}[t]
\includegraphics[width=7.5cm,angle=-90]{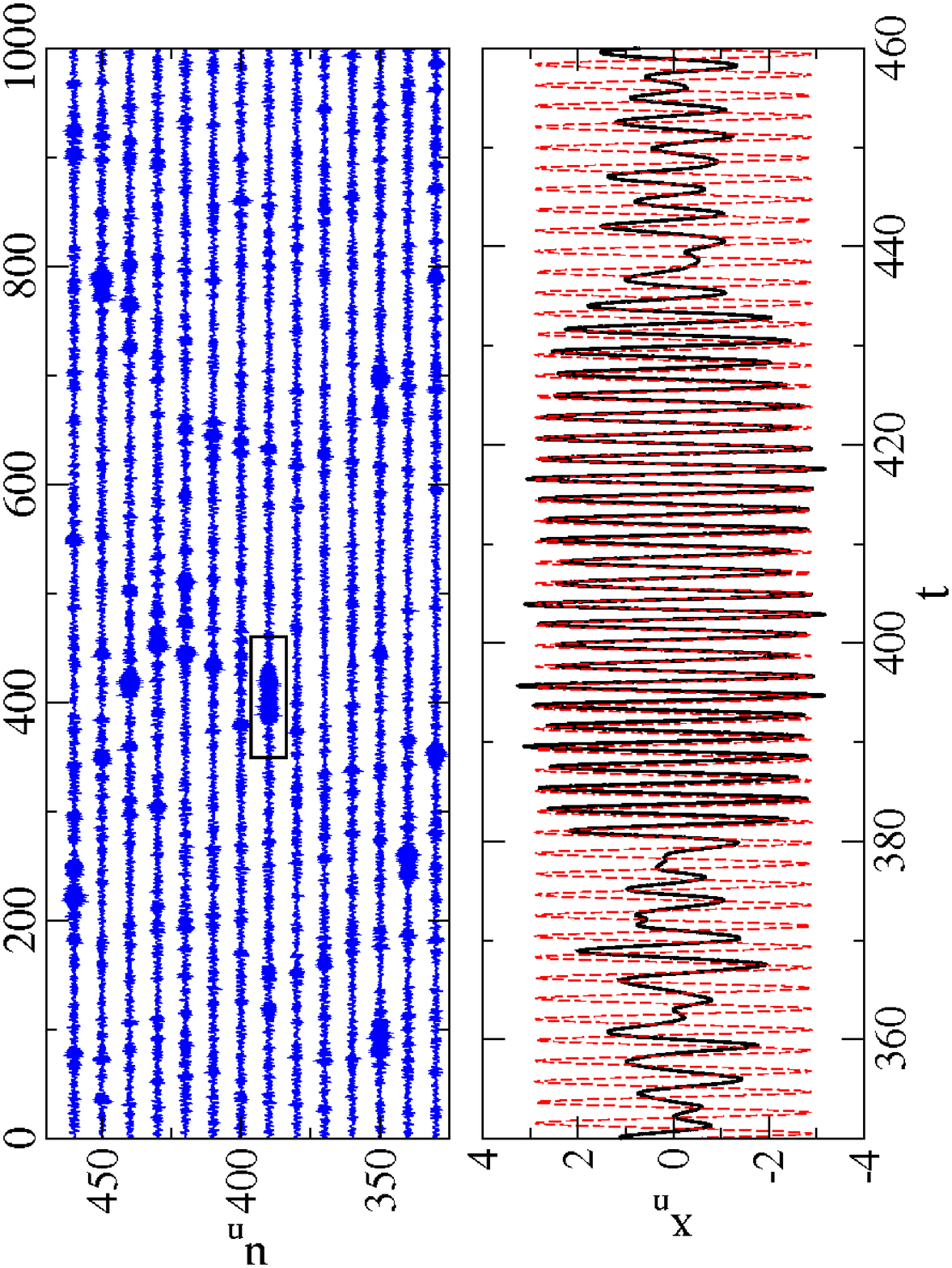}
\caption{
\label{fig:2}
(Color online) Same as in Fig.~\ref{fig:1} but for a subthreshold amplitude $F_0=0.6$ and noise strength $D=0.3$. The bottom panel shows a magnification of part of the trajectory $x_n(t)$ of one oscillator (boxed in the upper panel), as well as the time evolution of the center-oscillator of a deterministic ($D=0$) and dissipative breather ($\alpha=0.1$) with suprathreshold $F_0=1$ and same driving frequency (dashed-red line).
}
\end{figure}

Having summarized the results of the deterministic case, let us now 
turn to the analysis of the effect of thermal fluctuations and how 
they modify this picture. Starting with the suprathreshold case ($F_0>F_\mathrm{th}$), we would expect that the addition of noise provokes the spontaneous formation of breathers with the frequency dictated by the external driving, even if the system starts from an uniform configuration. This is indeed what is observed, as shown in Fig.~\ref{fig:1}, where a typical realization of the 
oscillators' trajectories has been plotted. The frequencies and amplitudes of the observed noise-induced breathers are very similar to the deterministic ST breather depicted in Fig.~\ref{fig:0}. The ST breather has the smallest size and the lowest energy of all possible collective modes at the frequency of the driver, thus being the most probable breather to be induced by noise. However, unlike the deterministic case, the breathers are not permanent, and the same random force which has generated them is able to destroy them after a while.

In the subthreshold case the driving force is not able by itself to sustain the breathers. Remarkably, however, the simulations show that these breathers are still produced in the system by the concerted action of noise and the driving force. This is shown in Fig.~\ref{fig:2}. The bottom panel shows a comparison of one of these noise-induced breathers with the deterministic stable breather obtained with a suprathreshold driving force.  This effect resembles one of the main features of the earliest manifestations of SR: the noise-induced hopping events of a Brownian particle in a bistable potential subject to a externally applied force which taken alone is not sufficient to produce the hopping events \cite{gamhan98}. Noise appears then as beneficial, enabling system transitions 
between the uniform background state and the coherent, localized 
breather patterns. On the other hand, we would expect that if the thermal fluctuations are very large, they will destroy any degree of order in the system, just like in the mentioned Brownian particle example, producing the non-monotonic behavior characteristic of the SR phenomenon. This scenario should offer a generic mechanism for the formation of breather-like excitations 
in realistic nonlinear lattices where fluctuations are present.

\begin{figure}[t]
\includegraphics[width=7.5cm,angle=0]{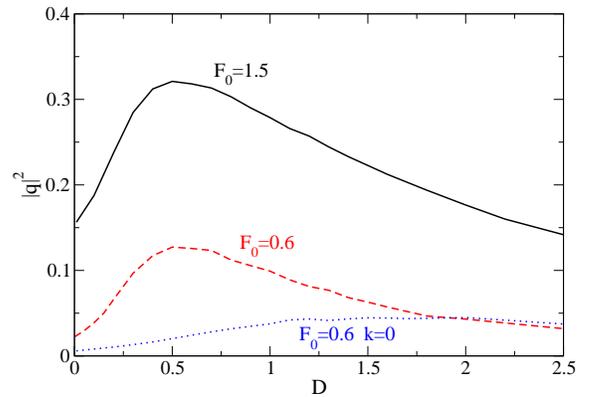}
\caption{
\label{fig:3}
(Color online) The Fourier amplitude $|q|^2$, defined in Eq.~(\ref{eq:qdef}), as a function of the noise strength $D$ for a suprathreshold driving $F_0=1.5$ (solid line), a subthreshold driving $F_0=0.6$ (dashed line), and for a system with $F_0=0.6$ but without interaction between the oscillators, i.e. $k=0$ (dotted line).
}
\end{figure}

In order to quantify these effects, we have computed numerically the Fourier coefficient of an arbitrary oscillator at the frequency of the external driver,
\begin{equation}
q=\frac{2}{T}\int_0^{2\pi/\Omega} dt\,e^{-i\Omega t}\langle x_1(t)\rangle_{\infty},
\label{eq:qdef}
\end{equation}
where the notation $\langle \ldots \rangle$ indicates an average over noise realizations and the subindex $\infty$ indicates the long time limit of the noise average, i.e., its value after waiting for time long enough for the transients to die out. Every time the oscillator $x_1$ is part of a breather, or a collective attempt to form a breather mode at the frequency of the driver, will result in 
a larger coefficient $q$. The choice of the oscillator in (\ref{eq:qdef}) is irrelevant because of the periodic boundary conditions. The simulations were performed with $N=40$ and $N=60$ oscillators, the larger value carried out in order to check that the results did not depend on the system size.

We present in Fig.~\ref{fig:3} the values of the amplitude $|q|^2$ as a function of the noise strength $D$ for a suprathreshold and subthreshold driving.
The non-monotonic behavior with noise, a recognized signature of the SR phenomenon, is very clear in both the suprathreshold and subthreshold situations. Nevertheless, let us note that a non-monotonic behavior is also obtained when we inhibit breather formation by canceling the interaction between the oscillators. The dotted line in Fig.~\ref{fig:3} shows this feature for a system with $k=0$. This is a well-known phenomenon for any  single-oscillator system subject to a nonlinear monostable potential, both in the underdamped \cite{stoste93} or overdamped \cite{evsrei04} regime. In the former, the peak is produced by a matching between the driving frequency and the intrinsic frequency associated with the most probable excitation energy, which is a monotonic function of the noise strength \cite{stoste93}. In fact, a simple classical mechanics analysis of the undriven Hamiltonian system ($\alpha,D, F_0=0$) with $k=0$ shows that a periodic solution with frequency $\omega=3$ corresponds to an energy per particle value of $E/N=36$. Under thermal equilibrium ($F_0=0$), we need to raise the temperature to a value 
such that the noise strength is $D=1.67$ in order to obtain an average energy per particle of the same value. It can be seen in Fig.~\ref{fig:3} that this value of $D$ is in good agreement with the position of the maximum. A similar argument also holds when we switch on the interaction between the oscillators, though now collective excitations associated with lower energies are allowed, resulting in larger amplitudes with a peak at a lower value of $D$. 
The total energy associated with the breather presented in Fig.~\ref{fig:0} is $E=30$. If we consider that roughly only $l=5$ oscillators are needed to form the ST breather, an average energy per particle of $E/N=6$ would be necessary in order to optimize breather formation. This value is obtained at thermal equilibrium when $D=0.7$, which is a reasonable estimation of the peak position observed in Fig.~\ref{fig:3}. A better estimation, namely a peak at $D=0.47$, is obtained if we choose $l=7$ oscillators. The appropriate value of $l$ is determined by the correlation length, as discussed below.

\begin{figure}[t]
\includegraphics[width=7.5cm,angle=0]{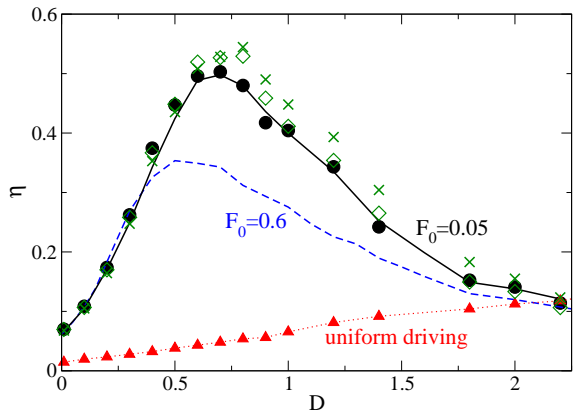}
\caption{
\label{fig:4}
(Color online) Spectral amplification $\eta=|q|^2/F_0^2$ as a function of noise. Filled circles depict the linear response results, computed numerically using Eq.~(\ref{eq:eta}), and the solid line the simulation results for $F_0=0.05$. The dashed line corresponds to $F_0=0.6$. Crosses and diamonds correspond to a linear response calculation neglecting correlations between oscillators separated by more than 2 and 3 sites, respectively. Filled triangles depict the linear response prediction for a system with a uniform external driving $F_n(t)=F_0\sin(\Omega t)$ (for all $n$) instead of (\ref{eq:force}). The dotted line is a guide to the eye.
}
\end{figure}

For an arbitrarily small driving amplitude $F_0$, linear response theory \cite{hantho82,Zwanzig01} predicts
\begin{equation}
 \langle x_1(t)\rangle_\infty=\int_0^\infty ds\, \chi(s) f(t-s),
\label{eq:linear}
\end{equation}
where $f(t)=F_0\sin(\Omega t)$ and $\chi(t)$ is the response function. This function can be expressed in terms of equilibrium time correlation functions,
\begin{equation}
 \chi(t)=- \frac{\alpha}{D}\frac{d}{dt}\langle x_1(t)M(0)\rangle_\mathrm{eq},
\label{eq:chi}
\end{equation}
where $\langle\ldots\rangle_\mathrm{eq}$ denotes equilibrium averages in the absence of driving and $M(t)=\sum_{n=1}^N (-1)^n x_n(t)$ is part of the perturbing Hamiltonian $-M f(t)$ associated with the external driving. The calculation of the Fourier coefficient of (\ref{eq:linear}), together with the definitions $\chi_r=\int_0^\infty \chi(t) \cos(\Omega t)$ and  $\chi_i=\int_0^\infty \chi(t) \sin(\Omega t)$, yields the following spectral amplification $\eta=|q|^2/F_0^2$
\begin{equation}
 \eta=(\chi(0)-\chi_i\Omega)^2+(\chi_r\Omega)^2.
\label{eq:eta}
\end{equation}
Fig.~\ref{fig:4} shows the linear response results obtained from simulations at thermal equilibrium. It is seen that a driving amplitude of $F_0=0.05$ is small enough to be within linear response theory. 
Due to the presence of noise and dissipation, the correlation between two separated oscillators decreases rapidly as their separation is increased. In fact, Fig.~\ref{fig:4} shows that a good approximation is obtained if the time correlation functions $\langle x_1(t)x_n(0)\rangle_\mathrm{eq}$ corresponding to oscillators separated by more than 2 sites ($|n-1|>2$) are neglected in Eq.~(\ref{eq:chi}). This approximation is consistent with the above energy matching argument with $l=5$, and the dominant dynamical role of ST breathers, leading to the peak estimation at $D=0.7$. Larger driving amplitudes are associated with larger correlation lengths, producing a peak at slightly lower values of $D$. A detailed
quantification of the dynamical role of multibreather excitations \cite{milkho09,kholep01,kholep02} would be a natural topic for future studies.

Finally, the calculated correlation functions allow us to readily compute the system response when the driving force acting on each oscillator is not staggered but spatially uniform, i.e. $M(t)=\sum_{n=1}^N x_n(t)$. The dotted line in Fig.~\ref{fig:4} shows that the spectral amplification is very similar to that when breather formation was strictly forbidden (the case $k=0$ shown in Fig.~\ref{fig:3}), lacking the stronger peak at lower noise levels. As expected from the deterministic analysis, this uniform driving force is very ineffective in boosting breather formation.

\paragraph{Summary.}
We have demonstrated the spontaneous formation of breathers in a hard $\phi^4$ lattice subject to a staggered driving force. The SR mechanism is different from the one observed in the paradigmatic Brownian particle in a bistable potential. Here, instead of a stochastic synchronization between noise-induced hopping events and the frequency of the external driving, the formation of noise-induced breathers with the selected driving frequency is optimized for a given level of fluctuations such that the average energy per particle matches the intrinsic energy of the breather mode. Further, this effect is shown to be suppressed if a spatially-uniform driving is used instead of a staggered one. We expect that the results reported in this paper may be useful for the generation and manipulation of breathers in many practical situations,
and to be independent of dimension and of the specific nature of the
nonlinearity, provided that the model is monostable.

\paragraph{Acknowledgments.}
This research was supported by the Ministerio de Ciencia e Innovaci\'on of Spain (FIS2008-02873 and FIS2008-04848), as well as by NSF-CAREER, NSF-DMS-0806762.
P.G.K. is grateful to D. Cai for originally suggesting this problem and
numerous discussions.


\begin{thebibliography}{23}
\expandafter\ifx\csname natexlab\endcsname\relax\def\natexlab#1{#1}\fi
\expandafter\ifx\csname bibnamefont\endcsname\relax
  \def\bibnamefont#1{#1}\fi
\expandafter\ifx\csname bibfnamefont\endcsname\relax
  \def\bibfnamefont#1{#1}\fi
\expandafter\ifx\csname citenamefont\endcsname\relax
  \def\citenamefont#1{#1}\fi
\expandafter\ifx\csname url\endcsname\relax
  \def\url#1{\texttt{#1}}\fi
\expandafter\ifx\csname urlprefix\endcsname\relax\def\urlprefix{URL }\fi
\providecommand{\bibinfo}[2]{#2}
\providecommand{\eprint}[2][]{\url{#2}}

\bibitem[{\citenamefont{Flach and Gorbach}(2008)}]{FG08}
\bibinfo{author}{\bibfnamefont{S.}~\bibnamefont{Flach}} \bibnamefont{and}
  \bibinfo{author}{\bibfnamefont{A.}~\bibnamefont{Gorbach}},
  \bibinfo{journal}{Phys. Rep.} \textbf{\bibinfo{volume}{267}},
  \bibinfo{pages}{1} (\bibinfo{year}{2008}).

\bibitem[{\citenamefont{MacKay and Aubry}(1994)}]{macaub94}
\bibinfo{author}{\bibfnamefont{R.}~\bibnamefont{MacKay}} \bibnamefont{and}
  \bibinfo{author}{\bibfnamefont{S.}~\bibnamefont{Aubry}},
  \bibinfo{journal}{Nonlinearity} \textbf{\bibinfo{volume}{7}},
  \bibinfo{pages}{1623} (\bibinfo{year}{1994}).

\bibitem[{\citenamefont{Campbell et~al.}(2004)\citenamefont{Campbell, Flach,
  and Kivshar}}]{cfk04}
\bibinfo{author}{\bibfnamefont{D.~K.} \bibnamefont{Campbell}},
  \bibinfo{author}{\bibfnamefont{S.}~\bibnamefont{Flach}}, \bibnamefont{and}
  \bibinfo{author}{\bibfnamefont{Y.~S.} \bibnamefont{Kivshar}},
  \bibinfo{journal}{Phys. Today} \textbf{\bibinfo{volume}{57}},
  \bibinfo{pages}{43} (\bibinfo{year}{2004}).

\bibitem[{\citenamefont{Gammaitoni et~al.}(1998)\citenamefont{Gammaitoni,
  H{\"a}nggi, Jung, and Marchesoni}}]{gamhan98}
\bibinfo{author}{\bibfnamefont{L.}~\bibnamefont{Gammaitoni}},
  \bibinfo{author}{\bibfnamefont{P.}~\bibnamefont{H{\"a}nggi}},
  \bibinfo{author}{\bibfnamefont{P.}~\bibnamefont{Jung}}, \bibnamefont{and}
  \bibinfo{author}{\bibfnamefont{F.}~\bibnamefont{Marchesoni}},
  \bibinfo{journal}{Rev. Mod. Phys.} \textbf{\bibinfo{volume}{70}},
  \bibinfo{pages}{223} (\bibinfo{year}{1998}).

\bibitem[{\citenamefont{Marchesoni et~al.}(1996)\citenamefont{Marchesoni,
  Gammaitoni, and Bulsara}}]{margam96}
\bibinfo{author}{\bibfnamefont{F.}~\bibnamefont{Marchesoni}},
  \bibinfo{author}{\bibfnamefont{L.}~\bibnamefont{Gammaitoni}},
  \bibnamefont{and} \bibinfo{author}{\bibfnamefont{A.~R.}
  \bibnamefont{Bulsara}}, \bibinfo{journal}{Phys. Rev. Lett.}
  \textbf{\bibinfo{volume}{76}}, \bibinfo{pages}{2609} (\bibinfo{year}{1996}).

\bibitem[{\citenamefont{Jung and Mayer-Kress}(1995)}]{junmay95}
\bibinfo{author}{\bibfnamefont{P.}~\bibnamefont{Jung}} \bibnamefont{and}
  \bibinfo{author}{\bibfnamefont{G.}~\bibnamefont{Mayer-Kress}},
  \bibinfo{journal}{Phys. Rev. Lett.} \textbf{\bibinfo{volume}{74}},
  \bibinfo{pages}{2130} (\bibinfo{year}{1995}).

\bibitem[{\citenamefont{Vilar and Rub{\'{\i}}}(1997)}]{vilrub97}
\bibinfo{author}{\bibfnamefont{J.~M.~G.} \bibnamefont{Vilar}} \bibnamefont{and}
  \bibinfo{author}{\bibfnamefont{J.~M.} \bibnamefont{Rub{\'{\i}}}},
  \bibinfo{journal}{Phys. Rev. Lett.} \textbf{\bibinfo{volume}{78}},
  \bibinfo{pages}{2886} (\bibinfo{year}{1997}).

\bibitem[{\citenamefont{Chen et~al.}(1996)\citenamefont{Chen, Aubry, and
  Tsironis}}]{cheaub96}
\bibinfo{author}{\bibfnamefont{D.}~\bibnamefont{Chen}},
  \bibinfo{author}{\bibfnamefont{S.}~\bibnamefont{Aubry}}, \bibnamefont{and}
  \bibinfo{author}{\bibfnamefont{G.~P.} \bibnamefont{Tsironis}},
  \bibinfo{journal}{Phys. Rev. Lett.} \textbf{\bibinfo{volume}{77}},
  \bibinfo{pages}{4776} (\bibinfo{year}{1996}).

\bibitem[{\citenamefont{Miloshevich et~al.}(2009)\citenamefont{Miloshevich,
  Khomeriki, and Ruffo}}]{milkho09}
\bibinfo{author}{\bibfnamefont{G.}~\bibnamefont{Miloshevich}},
  \bibinfo{author}{\bibfnamefont{R.}~\bibnamefont{Khomeriki}},
  \bibnamefont{and} \bibinfo{author}{\bibfnamefont{S.}~\bibnamefont{Ruffo}},
  \bibinfo{journal}{Phys. Rev. Lett.} \textbf{\bibinfo{volume}{102}},
  \bibinfo{pages}{020602} (\bibinfo{year}{2009}).

\bibitem[{\citenamefont{Johansson and Aubry}(2000)}]{johaub00}
\bibinfo{author}{\bibfnamefont{M.}~\bibnamefont{Johansson}} \bibnamefont{and}
  \bibinfo{author}{\bibfnamefont{S.}~\bibnamefont{Aubry}},
  \bibinfo{journal}{Phys. Rev. E} \textbf{\bibinfo{volume}{61}},
  \bibinfo{pages}{5864} (\bibinfo{year}{2000}).

\bibitem[{\citenamefont{Sievers and Takeno}(1988)}]{sietak88}
\bibinfo{author}{\bibfnamefont{A.~J.} \bibnamefont{Sievers}} \bibnamefont{and}
  \bibinfo{author}{\bibfnamefont{S.}~\bibnamefont{Takeno}},
  \bibinfo{journal}{Phys. Rev. Lett.} \textbf{\bibinfo{volume}{61}},
  \bibinfo{pages}{970} (\bibinfo{year}{1988}).

\bibitem[{\citenamefont{Khomeriki et~al.}(2001)\citenamefont{Khomeriki, Lepri,
  and Ruffo}}]{kholep01}
\bibinfo{author}{\bibfnamefont{R.}~\bibnamefont{Khomeriki}},
  \bibinfo{author}{\bibfnamefont{S.}~\bibnamefont{Lepri}}, \bibnamefont{and}
  \bibinfo{author}{\bibfnamefont{S.}~\bibnamefont{Ruffo}},
  \bibinfo{journal}{Phys. Rev. E} \textbf{\bibinfo{volume}{64}},
  \bibinfo{pages}{056606} (\bibinfo{year}{2001}).

\bibitem[{\citenamefont{Khomeriki et~al.}(2002)\citenamefont{Khomeriki, Lepri,
  and Ruffo}}]{kholep02}
\bibinfo{author}{\bibfnamefont{R.}~\bibnamefont{Khomeriki}},
  \bibinfo{author}{\bibfnamefont{S.}~\bibnamefont{Lepri}}, \bibnamefont{and}
  \bibinfo{author}{\bibfnamefont{S.}~\bibnamefont{Ruffo}},
  \bibinfo{journal}{Physica D} \textbf{\bibinfo{volume}{168-169}},
  \bibinfo{pages}{152} (\bibinfo{year}{2002}).

\bibitem[{\citenamefont{Anninos et~al.}(1991)\citenamefont{Anninos, Oliveira,
  and Matzner}}]{annoli91}
\bibinfo{author}{\bibfnamefont{P.}~\bibnamefont{Anninos}},
  \bibinfo{author}{\bibfnamefont{S.}~\bibnamefont{Oliveira}}, \bibnamefont{and}
  \bibinfo{author}{\bibfnamefont{R.~A.} \bibnamefont{Matzner}},
  \bibinfo{journal}{Phys. Rev. D} \textbf{\bibinfo{volume}{44}},
  \bibinfo{pages}{1147} (\bibinfo{year}{1991}).

\bibitem[{\citenamefont{Langer}(1990)}]{lan67}
\bibinfo{author}{\bibfnamefont{J.~S.} \bibnamefont{Langer}},
  \bibinfo{journal}{Ann. Phys.} \textbf{\bibinfo{volume}{41}},
  \bibinfo{pages}{108} (\bibinfo{year}{1990}).

\bibitem[{\citenamefont{Parisi}(1998)}]{parisi}
\bibinfo{author}{\bibfnamefont{G.}~\bibnamefont{Parisi}},
  \emph{\bibinfo{title}{Statistical field theory}} (\bibinfo{publisher}{Perseus
  Books Group}, \bibinfo{address}{New York}, \bibinfo{year}{1998}).

\bibitem[{\citenamefont{Mar\'{\i}n and Aubry}(1996)}]{MA96}
\bibinfo{author}{\bibfnamefont{J.~L.} \bibnamefont{Mar\'{\i}n}}
  \bibnamefont{and} \bibinfo{author}{\bibfnamefont{S.}~\bibnamefont{Aubry}},
  \bibinfo{journal}{Nonlinearity} \textbf{\bibinfo{volume}{9}},
  \bibinfo{pages}{1501} (\bibinfo{year}{1996}).

\bibitem[{\citenamefont{Mar\'{\i}n et~al.}(2001)\citenamefont{Mar\'{\i}n, Falo,
  Mart\'{\i}nez, and Flor\'{\i}a}}]{MFMF01}
\bibinfo{author}{\bibfnamefont{J.~L.} \bibnamefont{Mar\'{\i}n}},
  \bibinfo{author}{\bibfnamefont{F.}~\bibnamefont{Falo}},
  \bibinfo{author}{\bibfnamefont{P.~J.} \bibnamefont{Mart\'{\i}nez}},
  \bibnamefont{and} \bibinfo{author}{\bibfnamefont{L.~M.}
  \bibnamefont{Flor\'{\i}a}}, \bibinfo{journal}{Phys. Rev. E}
  \textbf{\bibinfo{volume}{63}}, \bibinfo{pages}{066603}
  (\bibinfo{year}{2001}).

\bibitem[{\citenamefont{Stocks et~al.}(1993)\citenamefont{Stocks, Stein, and
  McClintock}}]{stoste93}
\bibinfo{author}{\bibfnamefont{N.~G.} \bibnamefont{Stocks}},
  \bibinfo{author}{\bibfnamefont{N.~D.} \bibnamefont{Stein}}, \bibnamefont{and}
  \bibinfo{author}{\bibfnamefont{P.~V.~E.} \bibnamefont{McClintock}},
  \bibinfo{journal}{J. Phys. A} \textbf{\bibinfo{volume}{26}},
  \bibinfo{pages}{L385} (\bibinfo{year}{1993}).

\bibitem[{\citenamefont{Evstigneev et~al.}(2004)\citenamefont{Evstigneev,
  Reimann, Pankov, and Prince}}]{evsrei04}
\bibinfo{author}{\bibfnamefont{M.}~\bibnamefont{Evstigneev}},
  \bibinfo{author}{\bibfnamefont{P.}~\bibnamefont{Reimann}},
  \bibinfo{author}{\bibfnamefont{V.}~\bibnamefont{Pankov}}, \bibnamefont{and}
  \bibinfo{author}{\bibfnamefont{R.~H.} \bibnamefont{Prince}},
  \bibinfo{journal}{Europhys. Lett.} \textbf{\bibinfo{volume}{65}},
  \bibinfo{pages}{7} (\bibinfo{year}{2004}).

\bibitem[{\citenamefont{H\"anggi and Thomas}(1982)}]{hantho82}
\bibinfo{author}{\bibfnamefont{P.}~\bibnamefont{H\"anggi}} \bibnamefont{and}
  \bibinfo{author}{\bibfnamefont{H.}~\bibnamefont{Thomas}},
  \bibinfo{journal}{Phys. Rep.} \textbf{\bibinfo{volume}{88}},
  \bibinfo{pages}{207} (\bibinfo{year}{1982}).

\bibitem[{\citenamefont{Zwanzig}(2001)}]{Zwanzig01}
\bibinfo{author}{\bibfnamefont{R.}~\bibnamefont{Zwanzig}},
  \emph{\bibinfo{title}{Nonequilibrium Statistical Mechanics}}
  (\bibinfo{publisher}{Oxford University Press}, \bibinfo{address}{New York},
  \bibinfo{year}{2001}).

\bibitem[{\citenamefont{Sato et~al.}(2007)\citenamefont{Sato, Yasui, Kimura,
  Hikihara, and Sievers}}]{satyas07}
\bibinfo{author}{\bibfnamefont{M.}~\bibnamefont{Sato}},
  \bibinfo{author}{\bibfnamefont{S.}~\bibnamefont{Yasui}},
  \bibinfo{author}{\bibfnamefont{M.}~\bibnamefont{Kimura}},
  \bibinfo{author}{\bibfnamefont{T.}~\bibnamefont{Hikihara}}, \bibnamefont{and}
  \bibinfo{author}{\bibfnamefont{A.~J.} \bibnamefont{Sievers}},
  \bibinfo{journal}{Europhys. Lett.} \textbf{\bibinfo{volume}{80}},
  \bibinfo{pages}{30002} (\bibinfo{year}{2007}).

\end{thebibliography}


\end{document}